# Swift Highly Charged Ion Channelling

**Denis Dauvergne**

IPNL, Université de Lyon, Lyon, F-69003, France; Université Lyon 1 and IN2P3/CNRS, UMR 5822, 4 rue E. Fermi, F-69622 Villeurbanne Cedex, France.

d.dauvergne@ipnl.in2p3.fr

**Abstract**. We review recent experimental and theoretical progress made in the scope of swift highly charged ion channelling in crystals. The usefulness of such studies is their ability to yield impact parameter information on charge transfer processes, and also on some time related problems. We discuss the cooling and heating phenomena at MeV/u energies, results obtained with decelerated H-like ion beams at GSI and with ions having an excess of electrons at GANIL, the superdensity effect along atomic strings and Resonant Coherent Excitation.

## 1. Ion Channelling – Interaction with a dense electron gas

If an incident parallel ion beam has an orientation close to a major crystallographic direction of a crystal, then the ion flux is redistributed inside the target, below the surface: channelled ions undergo a series of correlated collisions at small angle along atomic strings or planes that repel them at large distances from the target atoms. The flux distribution of channelled positive particles is then characterized by a depletion of small impact parameters, and by a flux peaking near the centre of the axial or planar channels. This property has been widely used over the last 40 years to probe the structure of ordered matter, allowing lattice location of impurities, evidencing the nature of defects, providing information on surfaces and on strained epitaxial layers. It has also provided information on basic particle-matter interactions [1,2]. Here we mainly report on this latter aspect.

A useful description of channelling trajectories is obtained within the continuum potential approximation. The continuum potential $U(\vec{r})$ at a position $\vec{r}$ in the transverse plane is obtained by averaging the target atomic potential along the crystallographic direction. The transverse energy of an ion of charge $Q$, velocity $v$, momentum $p$, incident with an angle $\Psi$ relative to the axis (plane) is $E_\perp = QU(\vec{r}) + (vp/2)\Psi^2$. The value of the transverse energy is given by the incidence conditions ($\vec{r}_i, \Psi_i$). Then one can consider that a channelled ion is trapped into a 2D- (axial channelling) or 1D- (planar channelling) potential if $E_\perp$ is lower than a value close to the potential $QU(\rho)$, where $\rho$ is of the order of the thermal vibration extension of the crystal atoms around a string or a plane. This implies in particular that the incidence angle $\Psi_i$ must be smaller than a critical value $\Psi_c$ [3]. If one neglects energy loss, multiple angular scattering on electrons and nuclei, and charge exchange (under certain conditions), $E_\perp$ is conserved during the motion of a particle channelled inside a crystal. The transverse energy sets the accessible transverse space $A(E_\perp)$, in which the particle is confined.

Thus, well channelled ions with low $E_\perp$ explore restricted $A(E_\perp)$ and interact mainly with the loosely bound valence or conduction electrons. Their energy loss is reduced compared to unchannelled ions. Also the charge exchange regime is strongly influenced by the specific impact parameter distribution. As only small recoil momentum can be transferred to target atoms in large impact parameter collisions, processes like Mechanical Electron Capture (MEC), also called non-radiative capture) and Nuclear Impact Ionization (NII) are suppressed for channelled ions. Thus direct ion-electron interaction processes govern charge exchange, whereas they are often dominated during ion-atom collisions in ordinary matter. Figure 1 illustrates various charge exchange processes observable in crystal channelling. Ionization occurs via target electron impact (EII) mostly above the threshold $E_{ion}= (M_{ion}/m_e)E_B$, where $E_B$ is the initial electron binding energy. Electron capture may occur either radiatively (REC) or resonantly, by means of Resonant Transfer and Excitation (RTE).

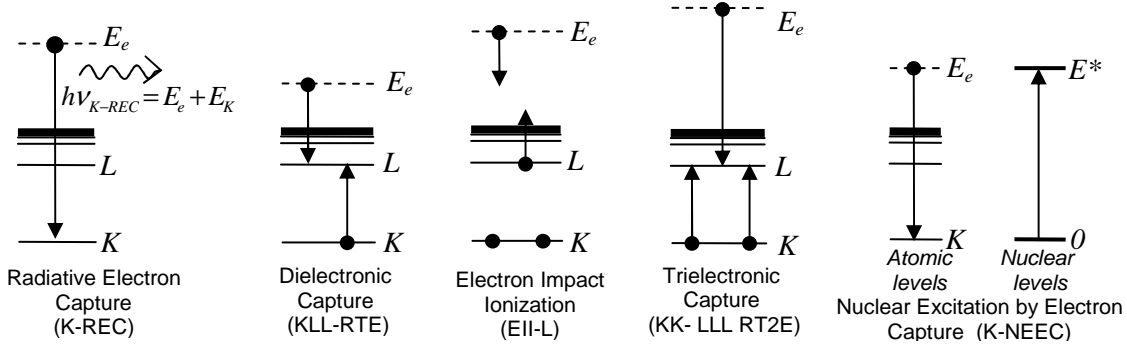

Figure 1: Schematic description, in the projectile frame, of charge exchange processes observable in highly charged ion channelling conditions. $E_e$ represents the kinetic energy of a target electron.

A review about charge exchange and electron excitation studies in aligned crystals by H.F. Krause and S. Datz [4] was published ten years ago; Resonant Coherent Excitation by the periodic crystal electric field was included. In the present review, we shall focus mainly on recent progress and results. In particular we will see in the following sections how local information (in the range 0.1-1 Å) can be extracted from charge exchange experiments, and how very short time scales (in the range of attoseconds to femtoseconds) may be involved in the evolution of excited states.

Before closing these preliminaries, let us emphasize that one can consider an aligned crystal as a very thick electron target. For instance, a few-µm thick Si crystal provides a $\sim 10^{20}$ e$^-$/cm$^2$ target, which makes channelling competitive with ion-electron collision experiments in storage rings or ion traps for investigating specific processes. An attempt was made at GANIL to observe the Resonant Trielectronic Recombination (RT2E, see Figure 1) with Kr$^{34+}$ ions [5]. An upper limit of the cross section in the mbarn range has been obtained. The resonant Nuclear Excitation by Electron Capture (NEEC) process is the time reverse of nuclear internal conversion [6]. This process has been reconsidered theoretically recently [7], and channelling could be the best method to measure it.

## 2. Breaking the reversibility rule at low energies

Channelling and blocking are reversed processes, in the sense that they provide identical angular dips: a channelling dip corresponds to the extinction of close collisions with target atoms for a beam parallel to a plane or an axis, and a blocking dip to the extinction of the ion flux along an axial or planar direction when ions are emitted isotropically from an atomic crystal site.

The reversibility rule implies that for a uniform angular distribution of the incident beam, the exit angular distribution should also be uniform, due to an exact compensation between the fractions of channelled and blocked particles inside the target. A violation of this reversibility rule was reported in 1999 by Assmann et al. [8], using ion energies in the MeV/u range.

As an example Figure 2 shows, for uniform angular distributions of the incident beam, the evolution with energy of the anisotropic angular distribution of Y ions emerging from a Si crystal. Clearly an enhancement of the flux is observed along each of the crystallographic directions at the highest energy (cooling), whereas flux disappears progressively with decreasing energy, depending on the direction, and, at the lowest energy, flux depletion along all directions is observed (heating).

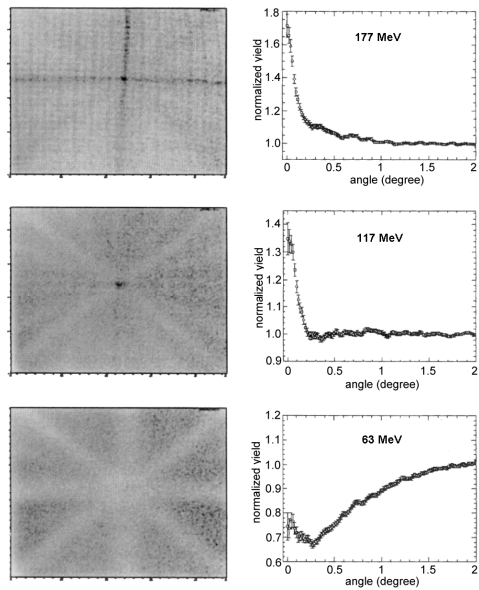

Figure 2: Flux distributions for Y ions transmitted through a 3.4 µm thick silicon crystal, for various energies at emergence. Left part: angular distributions recorded with track detectors. The central spot on the angular distribution corresponds to the <100> axis and the lines intersecting the spot correspond to the various planes intersecting the axis. Right part: circular averaged distributions around the <100> axis relative to the average level for a uniform distribution (data taken from ref. [9]).

Then the German collaboration performed a series of systematic studies by varying the projectile species, their energies, and the target species, thicknesses and orientations. A detailed report on their observations was published [9]. The transition from cooling to heating was studied as a function of the projectile atomic number and velocity, for silicon and nickel crystal targets. Among the various results presented, let us mention a few particularly interesting features:

- For projectiles with Z<22, only cooling is observed;
- Only cooling is observed along the <110> direction of silicon, which presents the broadest inter-atomic spacing in the transverse plane;
- for higher Z targets and projectiles, cooling and heating processes may coexist along the very same direction: for instance in the case of 40 MeV Ni ions, cooling is observed for the (100) plane of a Ni crystal at low transverse energy, whereas heating dominates at high transverse energy, and the overall effect is heating (see figure 9 of Ref. [9]).

As soon as they published their first observations, the authors explained most of the cooling and heating phenomenon by the impact parameter dependence of electron capture and loss probabilities by the projectiles [8, 10]. At such energies, charge exchange cross sections are very large, and charge equilibrium is reached rapidly in the target, at which the capture and loss rates become equal. If electron capture and loss occur at different mean impact parameters $r_c$ and $r_l$, then, after a cycle of $N$ captures and losses, the transverse energy of the ions travelling inside the crystal is changed by the amount $N[U(r_l) - U(r_c)]$ to first order, where $U$ is the continuum potential defined above. A more detailed theoretical description must account for the complex dependence of capture and loss probabilities on impact parameter. Grüner et al. have undertaken n-body, multi-collision Classical Transport Monte Carlo simulations (nN-CTMC) [9]. The first results of these simulations describe energy losses and charge state evolutions inside matter [11]. However, although this method is quite promising, a complete description of the cooling and heating features has not been obtained so far, due to the large computation power required. On the other hand, analytical calculations have been developed. For instance, C. Toepffer could calculate the transition energies observed in silicon by using scaling factors for capture and loss probabilities within the Born approximation [12].

Still some questions remain open. The main one is that there are no quantitative estimates of the cooling or heating strengths as a function of transverse energy, neither experimentally nor theoretically. Such studies would help for possible applications (could one use cooling for beam focusing?). The influence of cooling or heating on nuclear lifetime measurements by means of fission fragment blocking has to be evaluated. Finally, the link has to be made between the low energy case, where charge exchange takes place at large impact parameters with very large probabilities, and the higher energy case, with much smaller and more localized capture and loss probabilities, and where channelled ions can remain really frozen in their charge state. At one point, the influence of cooling at high energy will become negligible compared to the collisional heating (multiple scattering). However the upper energy limit for the influence of charge exchange is not yet known.

**3. Channelling as a local probe for charge exchange in the 10-100 MeV/u range**
Now we discuss ion channelling at higher energies, where no cooling and heating phenomena have been observed so far. In this high velocity regime ($v \gg v_0$) charge exchange processes involve deeply bound states of the projectile and/or the target, and thus precise information on their impact parameter dependence can be obtained.

3.1. Local Compton profile studies
The dominant electron capture processes for well channelled ions are REC and RTE, with cross sections of the order of the kbarn per target electron for highly charged ions in the present energy range. As noted previously, a thin crystal is a very dense electron target and, thus, these processes can be studied with very good statistics and used to characterize the electron gas sampled by channelled ions as a function of their transverse energy. Experiments can determine not only the electron density (*via* the capture probability) but also the longitudinal electron momentum distribution (Compton profile). Andersen et al. observed the KLL-RTE resonance profile with $Br^{33+}$ incident ions on a thin Si crystal as a function of incident energy, and for various selections of the total energy lost in axial channelling [13]. The resonance width is associated with the electron Compton profile. They observed a narrowing of the resonance profile for the best-channelled ions, in agreement with the local free electron gas approximation. In the same way, Andriamonje et al. measured X-rays corresponding to K-REC for 60 MeV/u $Kr^{36+}$ ions, both in axial and random orientations of a Si crystal [14]. Here also, the REC photon energy dispersion was dominated by the electron Compton profile. A clear impact parameter dependence of the Compton profile was deduced from the simulations fitting the data.

3.2. Charge exchange with decelerated highly charged ions
In this section we describe experiments performed with decelerated H-like heavy ions extracted from the ESR storage ring at GSI [15]. Channelling is then a unique tool to study in detail the interaction of highly charged ions, far from their charge equilibrium, with solid targets.

3.2.1. *Competition between MEC and REC.* For 20 MeV/u $U^{91+}$ ions colliding with a silicon target MEC cross sections are in the Mbarn range, and more than two orders of magnitude larger than REC ones. Thus, for an amorphous target, such incident ions capture many electrons by MEC, and reach charge equilibrium rapidly when the adiabaticity parameter $\eta_n = v_{ion}/v_n \approx 1$ for the last occupied shell (in this expression $v_n$ is the velocity of a bound electron in the *n*-shell). In channelling conditions, REC is found to be the dominant capture process for well-channelled ions, whereas it is prevented from occurring in random conditions, due to the rapid filling of inner-shells by MEC followed by cascades [16]. A quantitative analysis and simulations provided the absolute MEC and REC probabilities as a function of impact parameter [17]. In particular, MEC becomes the dominant recombination processes as ions approach atomic strings closer than 0.5 Å.

3.2.2. *Electron gas polarization.* 20 MeV/u $U^{91+}$ ions or 13 MeV/u $Pb^{81+}$ ions channelled in crystals induce a strong perturbation of the electron gas. The collective response (wake effect) induces a shift of the continuum energy level, which can be observed by means of radiative electron capture resulting from the transition from the continuum to a bound state of the projectile. REC photon energies are shifted consequently. The electron gas polarization also induces an increase of the local electron density. K- or L-REC by heavy ions are very local processes, in the sense that electrons are captured into very small radius orbitals. Thus the REC probability relative to the value in a non-perturbed electron gas should be enhanced by the local increase of the electron density. E. Testa et al. observed both effects [18]. As shown in Figure 3, the energy shift is in agreement with calculations using the linear response theory [19], and also with previous measurements by Tribedi et al., with lighter projectiles [20]. As for the local electron density, an enhancement of about 50% relative to theoretical values for K-REC has been observed. It is much smaller than predicted by the same linear response model. This shows that, for the small values of the adiabaticity parameter achieved in these experiments, the density fluctuations are distributed over quite large distances, rather than strongly localized in the vicinity of the heavy ions.

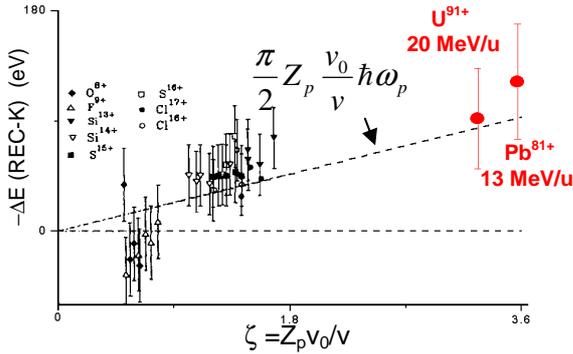

Figure 3: K- and L-REC line shift due to the wake potential induced by channelled heavy ions in silicon <110>. The data for $\zeta < 1.8$ are from Tribedi et al. [20]. The dashed line corresponds to linear response theory calculations [19] for a free electron gas of density 0.2 e$^-$/Å$^3$ ($\hbar\omega_p = 16.6\ eV$). The data for U$^{91+}$ and Pb$^{81+}$ ions are from Testa et al. [18].

3.2.3. *Deceleration of H-like ions.* As seen above, channelled ions with low transverse energy have a large probability to remain frozen in their incident charge state, even for decelerated H-like uranium ions far from their equilibrium charge state in solids. Although these ions experience a reduced electron density relative to random conditions, their high charge state may lead to very high-energy loss rates, mostly due to distant collisions with electrons. By extrapolating measurements for 20 MeV/u U$^{91+}$ ions, we show in Figure 4 that the energy loss of channelled frozen H-like ions may even exceed the energy loss of unchannelled ions that are only partially stripped of their electrons [2,16]. Since well-channelled ions may only capture electrons by REC, and the REC cross sections are not very large, one may transport and decelerate H-like ions inside relatively thick crystals. This has been confirmed recently by our collaboration and will be published soon. The questions following these experiments are i/ what is the maximum deceleration achievable without dechannelling and electron capture and ii/ what is the lowest energy at which an H-like U ion can be transported inside a crystal? So far, these two limits have not been reached. Eventually, one should reach the situation of section 2, where even channelled ions undergo many charge exchanges.

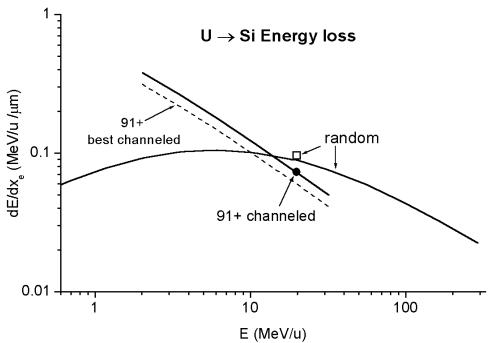

Figure 4: Expected evolution of the energy loss for channelled U$^{91+}$ ions in the <110> axis of silicon. The experimental data (circle: most probable energy loss in channelling, square: loss for random orientation) are taken from ref. [16]. The random energy loss curve corresponds to SRIM calculations [21]. The dashed curve corresponds to the loss experienced by the best-channelled ions.

## 4. Channelling as a time probe

4.1. Superdensity effect for high transverse energy ions

In this section we consider ions that have a transverse energy close to the critical value in axial channelling conditions instead of well-channelled ions. As long as these ions are not scattered at large angle during a single collision at very small impact parameter, they keep a channelling-like trajectory, *i.e.* parallel to the strings at the minimum distance of approach and they may spend a significant part of their time close to atomic strings. For incidence parallel to the axis, glancing collisions occur at the entrance of the crystal, and lead to the so-called surface peak in Rutherford scattering measurements.

These particular trajectories are characterized by several time scales governing the interaction of such ions with matter. These times are the following, for ions with 10 - 100 MeV/u kinetic energies:
- The time $T_{collis}$ between two collisions on successive atoms, with the same impact parameter (within ~0.1 Å due to the thermal vibrations) has a fixed value in the range $3.10^{-18}$ and $10^{-17}$ s.
- The length over which the trajectory remains tangent to an atomic string is of the order of 10 nm. This represents a time $T_{glancing}$ in the $10^{-16}$ s range.
- After being repelled by a string, ions will reach other strings after variable path lengths (in contrast with planar channelling, for which trajectories oscillate regularly). Moreover, the next string may be approached at a quite different minimum distance. However, let us consider that this time of flight $T_{free}$ in a medium free of target atomic cores is of the order of $10^{-15}$ s.
- Finally the time of flight inside a 1-10 µm thick crystal is in the $10^{-14}$-$10^{-12}$ s range.

The consequence of such a time distribution of the close collision rate is a very particular evolution of the electronic configuration of highly charged ions inside the crystal. On the one hand, excitation-ionization processes are favored close to the strings, since $T_{coll}$ and $T_{glancing}$ are shorter than most of the excited lifetimes, even for highly probable transitions. On the other hand, MEC is also enhanced at close impact parameters along atomic strings. However, for highly charged ions at moderate velocities, MEC is likely to occur into excited states, and the probability for the projectile to lose such electrons during further collisions is high, since decay into deeply bound states may occur at times of the order of $T_{free}$ or higher. Once the ions move away from the strings, both capture and ionization rates decrease, and radiative or auto-ionizing decay of the few surviving excited electrons may take place, at least partially. Thus, during the full traversal of the target, the ratio of probabilities for overall ionization and effective capture into stable states is larger for ions with a high transverse energy than in random conditions, when the time between close collisions is more uniformly distributed. This superdensity effect was described by A. L'Hoir et al. [22]. Experimental evidence was obtained with 29 MeV/u $Pb^{56+}$ ions incident on a thin silicon crystal. Near axial crystal orientation, a significant part of the charge state distribution, corresponding to the highest transverse energy ions, extends beyond the distribution for a random orientation toward higher charge states. This shows that thin crystals can be used to produce very highly stripped ions (in that case, up to He-like Pb ions). The authors of [22] described also how H-like incident uranium ions decelerated at lower energies are prevented from dressing with electrons at high transverse energy. A last remark, concerning glancing collisions along atomic strings, is that the energy loss rate is locally enhanced due to close collisions with target core electrons. If one extrapolates the values measured with MeV protons [23], the energy deposited during times of the order of $T_{glancing}$ may exceed 100 keV/nm for highly charged ions. Such values are above the maximum stopping power of amorphous matter for single projectiles. This opens interesting perspectives for the dynamics of damage induced by ions in matter.

4.2. Resonant Coherent Excitation
Okorokov [24] predicted in 1965 that channelled particles should be sensitive to the periodicity of the charge distribution along crystallographic directions, i.e. to harmonics of the string potential, U($\vec{r}$,z), the variation of which with z is now considered. The zero-order continuum potential is obtained after averaging over all the atoms of a string, as described in section 1. Higher order terms correspond to the maxima of the Fourier transform of U($\vec{r}$,z). Resonant Coherent Excitation (RCE) can occur when an excitation frequency $\Delta E/h$ of the projectile matches a harmonic of the collision frequency along an atomic string (1D-RCE): $\Delta E/h = k\gamma v/d$, where *v* is the projectile velocity, *d* the interatomic spacing along the string, $\gamma$ the Lorenz factor and $k$ a positive integer. Thus RCE is the resonant absorption (or emission) of a momentum corresponding to the reciprocal lattice vector, $\hbar G_k = k\, 2\pi\hbar/d$.

Under planar channelling conditions, one may adjust the excitation frequency by varying the angle $\theta$ relative to the atomic strings belonging to the plane [25]. For instance let us consider for simplicity that the plane is defined by two perpendicular axes, with inter-atomic distances $d_1$ and $d_2$. The 2D-RCE resonance condition becomes $\Delta E/h = \gamma v(k\cos\theta/d_1 + l\sin\theta/d_2)$, ($k$, $l$ positive integers).

Very recently a Japanese collaboration showed the possibility of performing 3D-RCE [26], by using the 3-fold periodicity in the direct space. Channelling is no longer considered, as ions cross planes of atoms periodically. By introducing a second angle $\Phi$ relative to the previous plane, and the third interatomic distance $d_3$ in the perpendicular direction, the resonance condition becomes
$\Delta E/h = \gamma v(k\cos\theta\cos\Phi/d_1 + l\sin\theta\cos\Phi/d_2 + m\sin\Phi/d_3)$, where $k$, $l$, $m$ are positive integers.

A review of the pioneering RCE experiments appeared in ref. [4]. In these experiments, the authors observed either a decrease of the survival fraction of frozen non-fully stripped incident ions, or an enhancement of K-X rays, corresponding to a resonance for n=1 to n=2 excitation. The ionization increase was due to the much larger ionization probability for n=2 states relative to that for n=1 states. Further complementary studies have been performed since then. Komaki *et al.* have extended these measurements in planar channelling to higher Z ions at higher energies [27]. They recorded simultaneously the energy loss by using a thin crystal target as an active silicon detector, so that they could obtain detailed information on the Stark splitting of the (n=2)-state of $Ar^{17+}$ ions as a function of the impact parameter in planar channelling conditions. Actually, the static transverse electric field sampled by an ion in the vicinity of an atomic plane (or row) is responsible for this Stark splitting. The stronger the electric field, the stronger the splitting. As ions of different transverse energies sample different accessible transverse spaces, the resonance shape was found to depend on the energy loss. This collaboration reported also on the enhancement of the production of convoy electrons due to RCE, followed by electron loss to the continuum [28] and on the resonant excitation of optically forbidden transitions from 2s to 3s and 3d sublevels of Li-like ions [29]. By means of anisotropic X-ray angular distributions, Azuma et al. recently demonstrated that RCE under planar channelling conditions populates aligned excited states [30]. The very recent discovery of 3D-RCE opens a new field for fundamental atomic physics [26]. In particular, the three components of the periodic electric field can be used to excite simultaneously several transitions, as in pump-probe laser experiments. Without channelling conditions, the absence of the slowly varying static electric field, responsible for the Stark broadening of the resonances, may improve the spectroscopic accuracy of these experiments.

In future experiments devoted to precision spectroscopy of highly charged ions, the ultimate physical limitation of the precision of RCE may come from the crystal itself, through experimental uncertainty in the crystal parameter (the interatomic distances $d_i$ in the above formulae).

To end this review on recent progress about RCE, let us mention a theoretical work by Balashov et al. [31] which discusses the possibility of using RCE in a crystal to populate metastable states of a projectile efficiently. Finally, theoretical descriptions and predictions of nuclear RCE have been published [32], but this process has not been observed yet.

## 5. Conclusion

As some questions and conclusions have already been given at the end of each section of this paper, let us give some perspectives on possible future studies with highly charged ions.

Decelerated H-like ion beams offer the opportunity to elucidate new physics, in energy and perturbation ranges that have not been completely explored. Interesting applications arise for atomic, surface science and nuclear physics. This should be part of the FLAIR program at the future FAIR facility (Darmstadt), where extracted H-like heavy ions below 10 MeV/u will be available.

The structured sequences of collisions along strings or planes for ions with relatively high transverse energy, as well as the pure electron gas target sampled by well channelled ions, provide original conditions for the study of the evolution of excited states inside a solid. Moreover, channelling allows populating given electronic states (e.g. 2s-REC) selectively. This is a challenging method for the detailed study of collisional or Stark mixing of metastable excited states [31, 33, 34].

The recent discovery of 3D-RCE opens new perspectives for fundamental atomic physics. It may also point out the need to account for coherence effects in determining excitation and ionization probabilities in solids. Indeed, solid targets are ordered on a nanometre scale, so interferences could either enhance or decrease transition probabilities as compared to independent collisions.

Improved precision atomic spectroscopy experiments are expected by increasing the beam energies, which is part of the SPARC letter of intent for FAIR. There, the very high energies achievable (~10 GeV/u for uranium ions) could also allow the study of nuclear RCE.

C. Cohen, A. L'Hoir, J.-C. Poizat and E. Testa are greatly acknowledged for their help in the preparation of this manuscript.